\documentclass{aa}

\usepackage{graphicx}
\usepackage{txfonts}
\usepackage{natbib}     
\usepackage{xcolor} 
\usepackage{threeparttable} 

\bibpunct{(}{)}{;}{a}{}{,} 

\usepackage{hyperref}
\begin{document}

   \title{CRIRES$^+$ detection of CO emissions lines and temperature inversions on the dayside of WASP-18b and WASP-76b}
 
   \author{F. Yan\inst{1,2}
                  \and
         L.~Nortmann\inst{2}
          \and
                 A. Reiners\inst{2}
                \and 
                N. Piskunov\inst{3}
                \and
                A. Hatzes\inst{4}
                \and
                U. Seemann\inst{5,2}
                \and       
                D. Shulyak\inst{6}
                  \and
                A. Lavail\inst{3}
                \and    
                A.~D.~Rains\inst{3}
                \and
                D.~Cont\inst{2}
                \and 
                M. Rengel\inst{7}
                \and
                F. Lesjak\inst{2}
                \and
                E. Nagel\inst{8,2,4}
                \and
                O. Kochukhov\inst{3}            
                \and
                S. Czesla\inst{4,8}
                \and
                L. Boldt-Christmas\inst{3}
                \and
                U. Heiter\inst{3}
                \and
                J. V. Smoker\inst{9,10}
                \and            
                F. Rodler\inst{9}
                \and            
                P. Bristow\inst{5}
                \and            
                R. J. Dorn\inst{5}
                \and
                Y. Jung\inst{5}
                \and
                T. Marquart\inst{3}
                \and
                E. Stempels\inst{3}
                }
  \institute{Department of Astronomy, University of Science and Technology of China, Hefei 230026, China\\
        \email{yanfei@ustc.edu.cn}
\and
Institut f\"ur Astrophysik und Geophysik, Georg-August-Universit\"at, Friedrich-Hund-Platz 1, D-37077 G\"ottingen, Germany
\and
Department of Physics and Astronomy, Uppsala University, Box 516, 75120 Uppsala, Sweden
\and
Th{\"u}ringer Landessternwarte Tautenburg, Sternwarte 5, 07778 Tautenburg, Germany
\and
European Southern Observatory, Karl-Schwarzschild-Str. 2, 85748 Garching bei M\"unchen, Germany
\and
Instituto de Astrof\'{\i}sica de Andaluc\'{\i}a - CSIC, Glorieta de la Astronom\'{\i}a s/n, 18008 Granada, Spain
\and
Max-Planck-Institute f\"ur Sonnensystemforschung, Justus-von-Liebig-Weg 3, 37077 G\"ottingen, Germany
\and
Hamburger Sternwarte, Universit{\"a}t Hamburg, Gojenbergsweg 112, 21029 Hamburg, Germany
\and
European Southern Observatory, Alonso de Cordova 3107, Vitacura, Santiago, Chile
\and
UK Astronomy Technology Centre, Royal Observatory, Blackford Hill, Edinburgh EH9 3HJ, UK
}
        \date{Received November 4, 2022; accepted February 8, 2023}


  \abstract
 {The dayside atmospheres of ultra-hot Jupiters (UHJs) are predicted to possess temperature inversion layers with extremely high temperatures at high altitudes. We observed the dayside thermal emission spectra of WASP-18b and WASP-76b with the new CRIRES$^+$ high-resolution spectrograph at near-infrared wavelengths. Using the cross-correlation technique, we detected strong CO emission lines in both planets, which confirms the existence of temperature inversions on their dayside hemispheres. 
The two planets are the first UHJs orbiting F-type stars with CO emission lines detected; previous detections were mostly for UHJs orbiting A-type stars.
Evidence of weak $\mathrm{H_2O}$ emission signals is also found for both planets. We further applied forward-model retrievals on the detected CO lines and retrieved the temperature-pressure profiles along with the CO volume mixing ratios. The retrieved logarithmic CO mixing ratio of WASP-18b ($-2.2_{-1.5}^{+1.4}$) is slightly higher than the value predicted by the self-consistent model assuming solar abundance. For WASP-76b, the retrieved CO mixing ratio ($-3.6_{-1.6}^{+1.8}$) is broadly consistent with the value of solar abundance. In addition, we included the equatorial rotation velocity ($\varv_\mathrm{eq}$) in the retrieval when analyzing the line profile broadening. The obtained $\varv_\mathrm{eq}$ is $7.0\pm{2.9}$ km\,s$^{-1}$ for WASP-18b and $5.2_{-3.0}^{+2.5}$ km\,s$^{-1}$ for WASP-76b, which are consistent with the tidally locked rotational velocities.
 }
   \maketitle

%

\section{Introduction}
Atmospheric temperature inversions have been predicted to exist on the daysides of hot gas giants due to the absorption of strong stellar irradiation in the ultraviolet and visible wavelengths by species such as iron, hydrogen, and TiO/VO \citep[e.g.,][]{Hubeny2003, Fortney2008, Lothringer2018, Garcia2019, Fossati2021-NLTE}. However, early searches for inversion layers in intermediate-hot planets were not successful \citep[e.g.,][]{Hansen2014, Schwarz2015}. Past observations (up to a few years ago) revealed the presence of temperature inversions on the dayside hemisphere of several ultra-hot Jupiters (UHJs), which are gas giants with extremely high equilibrium temperatures ($T_{\mathrm{eq}}$ > 2000\,K).
 
The discovery of temperature inversions is normally achieved by detecting spectral features in emission  with either high- or low-resolution spectroscopy.
For example, atomic iron (\ion{Fe}{i}) emission lines have been detected with high-resolution spectroscopy in KELT-9b \citep{Pino2020, Kasper2021}, WASP-189b \citep{Yan2020}, WASP-33b \citep{Nugroho2020W33, Cont2021, Herman2022}, and KELT-20b \citep{Yan2022-KELT20b, Borsa2022-KELT20b, Johnson2022, Kasper2023}; carbon monoxide (CO) emission lines have been detected in WASP-189b, WASP-33b, and MASCARA-1b \citep{Yan2022-GIANO, vanSluijs2022, Holmberg2022}. A list of discovered chemical species with high-resolution emission observations is presented in Table~\ref{tab-UHJs}. Low-resolution observations have also revealed the existence of inversion layers by detecting the $\mathrm{H_2O}$ and CO emission bands with the \textit{Hubble} and \textit{Spitzer} space telescopes \citep[e.g.,][]{Evans2017, Mansfield2021, Fu2022}.
In addition to emission spectroscopy, a variety of chemical species has been found at the terminators of UHJs through transmission spectroscopy, including hydrogen, \ion{Fe}{i}, \ion{Fe}{ii}, \ion{Ca}{ii}, \ion{O}{i}, and TiO \citep{Yan2018, Hoeijmakers2018, Casasayas-Barris2018, Yan2019, Sing2019, Cabot2020, Bello-Arufe2022, Stangret2022, Prinoth2022, Borsa2022}.

\begin{table*}
\small 
\caption{Summary of UHJ observations with high-resolution emission spectroscopy.}             
\label{tab-UHJs}      
\centering                          
\begin{threeparttable}
        \begin{tabular}{l c c c c}        
        \hline\hline \noalign{\smallskip}                 
                Planet & $T_\mathrm{eq}$ (planet) & $T_\mathrm{eff}$ (star)  &  Discovered species on dayside & References\\     
        \hline     \noalign{\smallskip}                   
    \rule{0pt}{2.5ex}KELT-9b & 3921\,K & 9600\,K & \ion{Fe}{i} & [1][2][3]\\  
    \rule{0pt}{2.5ex}KELT-20b/MASCARA-2b & 2261\,K & 8730\,K & \ion{Fe}{i}, \ion{Si}{i}, \ion{Fe}{ii}, \ion{Cr}{i}, \ion{Ni}{i} & [4][5][6][7][8]\\     
    \rule{0pt}{2.5ex}WASP-189b & 2641\,K & 8000\,K & \ion{Fe}{i}, CO & [9][10]\\ 
    \rule{0pt}{2.5ex}MASCARA-1b & 2594\,K & 7554\,K & CO, $\mathrm{H_2O}$ & [11]\\    
    \rule{0pt}{2.5ex}WASP-33b & 2710\,K & 7430\,K & \ion{Fe}{i}, \ion{Si}{i}, \ion{V}{i}, \ion{Ti}{i}, CO, OH, TiO & [5][10][12][13][14][15][16][17]\\  
    \rule{0pt}{2.5ex}WASP-18b & 2411\,K & 6400\,K & CO, $\mathrm{H_2O}$, OH & this work, [18] \\
    \rule{0pt}{2.5ex}WASP-76b & 2228\,K & 6330\,K & CO, $\mathrm{H_2O}$ & this work \\
\noalign{\smallskip}  \hline                                   
        \end{tabular}
\tablefoot{
  \tablefoottext{1}{\cite{Pino2020}.}
  \tablefoottext{2}{\cite{Kasper2021}.}
  \tablefoottext{3}{\cite{Pino2022}.}
  \tablefoottext{4}{\cite{Yan2022-KELT20b}.}
  \tablefoottext{5}{\cite{Cont2022}.}
  \tablefoottext{6}{\cite{Borsa2022-KELT20b}.}
  \tablefoottext{7}{\cite{Johnson2022}.}
  \tablefoottext{8}{\cite{Kasper2023}.}
  \tablefoottext{9}{\cite{Yan2020}.}  
  \tablefoottext{10}{\cite{Yan2022-GIANO}.}
  \tablefoottext{11}{\cite{Holmberg2022}.}   
  \tablefoottext{12}{\cite{Nugroho2020W33}.}   
  \tablefoottext{13}{\cite{Cont2021}.} 
  \tablefoottext{14}{\cite{Nugroho2021}.}
  \tablefoottext{15}{\cite{Herman2022}.}
  \tablefoottext{16}{\cite{Cont2022b}.} 
  \tablefoottext{17}{\cite{vanSluijs2022}.} 
  \tablefoottext{18}{\cite{Brogi2022}.} 
}        
\end{threeparttable}      
\end{table*}
%

In this paper we present the discovery of CO emission lines and evidence of $\mathrm{H_2O}$ signals in \object{WASP-18b} and \object{WASP-76b} with the CRIRES$^+$ instrument (the CRyogenic InfraRed Echelle Spectrograph upgrade project), a recently upgraded near-infrared high-resolution spectrograph on the Very Large Telescope. Both WASP-18b and WASP-76b are UHJs orbiting F-type stars. Previous thermal emission observations from the \textit{Hubble} and \textit{Spitzer} telescopes have shown evidence of temperature inversions on the two planets \citep{Sheppard2017, Arcangeli2018, Edwards2020}.
The transmission spectrum of WASP-76b has been extensively studied, and asymmetric spectral features from the two terminator limbs have been observed \citep[e.g.,][]{Ehrenreich2020, Tabernero2021, Casasayas-Barris2021, Sanchez-Lopez2022}. However, CO has not been detected in its transmission spectrum yet. There is no detection of WASP-18b's transmission spectra in the literature, probably because the planet has a large surface gravity that hinders the probing of transmission signals.
Therefore, our detection of CO emission lines not only confirms the existence of inversion layers in the two planets but also provides a unique diagnostic for studying their atmospheric properties. We note that, in parallel to this work, a study of WASP-18b's emission spectrum has been presented by \cite{Brogi2022}.

The manuscript is organized as follows. In Sect. 2 we describe the observations and data reduction procedures. In Sect. 3 we describe the method for detecting CO and $\mathrm{H_2O}$ emission lines.
In Sect. 4 we present the detection results and the atmosphere retrievals, along with the discussion. Conclusions are presented in Sect. 5.

\section{Observations and data reduction}
As part of guaranteed time observations, we observed WASP-18b on 9 October 2021 and WASP-76b on 31 October 2021 with the CRIRES$^+$ instrument. Compared to the original CRIRES, the CRIRES$^+$ upgrade transformed the instrument into a cross-dispersed spectrograph with a large simultaneous wavelength coverage \citep{Dorn2023}.
The instrument started its regular science observations after the science verification in September 2021 \citep{Leibundgut2022}. WASP-18b and WASP-76b are the first two exoplanets that have been observed by the CRIRES$^+$ consortium for atmosphere studies.

The observations of the two planets were performed at orbital phases before or after the secondary eclipses when their dayside hemispheres were visible (see Table \ref{obs_log} for the phase coverage and a summary of the observations). We employed the nodding mode, in which the target is observed at two positions (A and B) along the slit in order to optimally remove sky emission and detector artifacts from each frame. We used the K2148 wavelength setting (1972--2452\,nm) with a detector integration time (DIT) of 120\,s. Metrology was employed to improve the wavelength calibration.
We chose the 0.2\,\arcsec slit to reach a high spectral resolution and employed the adaptive optics (AO) system.
The observation of WASP-18b lasted for $\sim$ 3.8\,hr; however, the flux level dropped by half after the first 1.2\,hr due to bad seeing conditions. The observation of WASP-76b lasted for $\sim$ 1.7\,hr under good weather conditions. The signal-to-noise ratio (S/N) per pixel around 2.43\,$\mathrm{\mu}$m is 15$\sim$60 for the WASP-18b observation and 35$\sim$55 for the WASP-76b observation.
The resolving power of the spectrograph ($R$) is around 86\,000 -- 110\,000 for full slit illumination when using the 0.2\,\arcsec slit. The resolution can be even higher with good AO corrections under ideal seeing conditions. We measured the resolution of the observed spectra by calculating the instrumental broadening profile of the telluric absorption lines. For the WASP-18b spectra, the resolution is $\sim$ 120\,000 at the beginning of the observation and drops to $\sim$ 98\,000 under bad seeing conditions. For WASP-76b spectra, the resolution is $\sim$ 130\,000. In this work, we adopted $R\sim$ 120\,000 for WASP-18b data and $R\sim$ 130\,000 for WASP-76b.

\begin{figure*}
   \centering
   \includegraphics[width=0.7\textwidth]{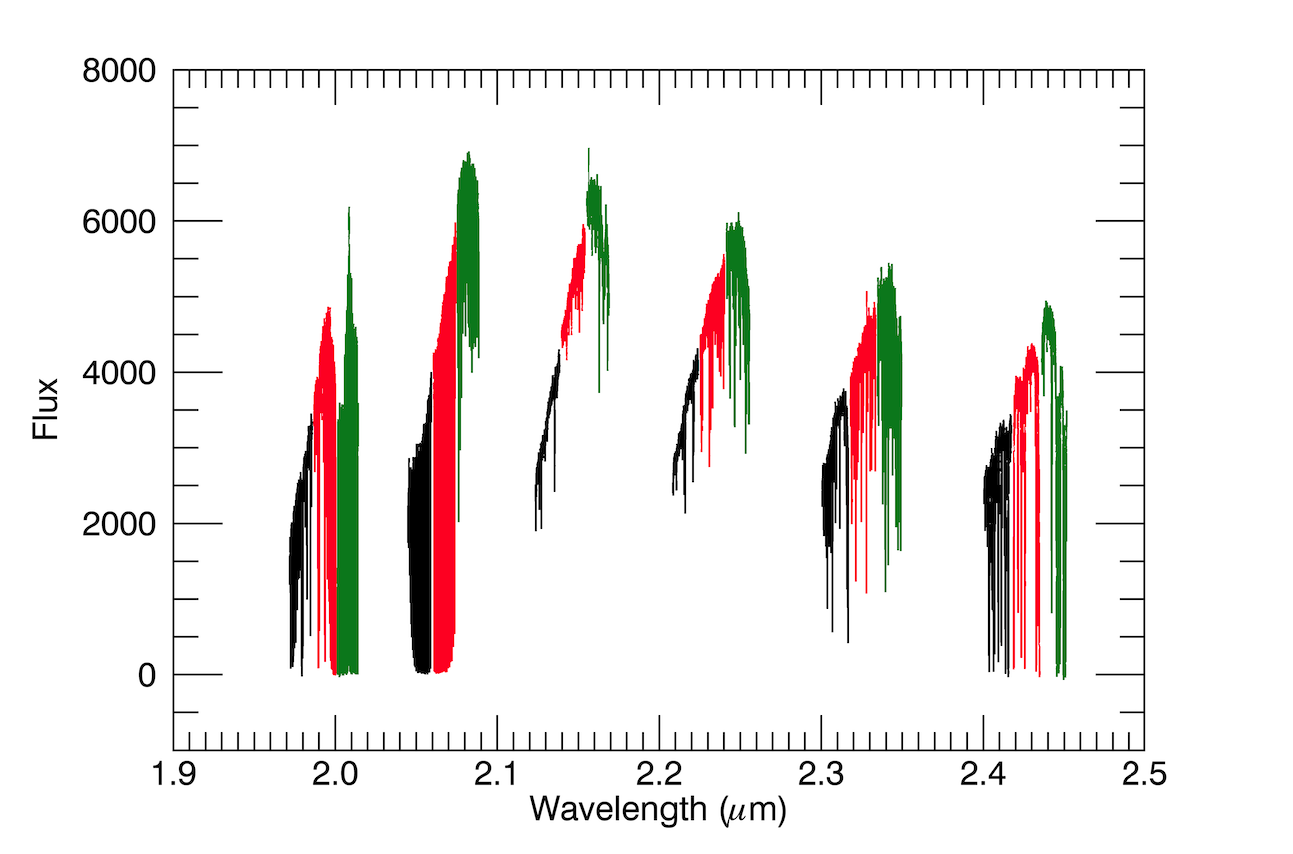}
      \caption{Example of the reduced CRIRES$^+$ spectrum of WASP-76 using the K2148 wavelength setting. The colors represent the three signal detectors.}
         \label{spec-demo}
   \end{figure*}

We reduced the raw spectra using the ESO CRIRES$^+$ pipeline (version 1.1.4) with the \texttt{EsoRex} tool. This version of the pipeline significantly improves the accuracy of the wavelength calibration compared to previous versions.
To achieve a high spectral S/N, we used deep flats taken on 4 October 2021 for flat fielding. Deep flats are taken regularly by the instrument team with NDIT (number of DIT) = 50 (i.e., 50 exposures for each flat frame). The nodding pairs were reduced using the \texttt{cr2res\underline{~}obs\underline{~}nodding} command. We set the slit-function oversampling factor to 12 and turned on the \texttt{subtract\underline{~}nolight\underline{~}rows} during the reduction. The pipeline provides the extracted spectra at positions A and B separately. The A and B spectra were treated as independent data sets in the subsequent reductions.
Each spectrum consists of six spectral orders with three segments per order, which correspond to the three chips of the detector. We trimmed the first and last 20 pixels of each spectral segment because the data points there typically have low qualities. We performed the normalization on each spectral segment and then merged all 18 segments into one spectrum (Fig.~\ref{spec-demo}).

To check the accuracy of the wavelength calibration, we compared the wavelength solution provided by the pipeline with a spectral template of telluric absorption lines. We find that the wavelength calibration is accurate at the K2148 setting with an uncertainty below 1 km\,s$^{-1}$. Therefore, we did not apply any additional refinement to the wavelength calibration. We also checked the relative stability of the spectrograph between consecutive exposures by computing the drifts of telluric lines between the spectral frames. The calculated drifts are below 0.4 pixels during the observation, indicating that the spectrograph is relatively stable. We further corrected the drifts to align the spectra.

%
\begin{table*}
\small
\caption{Observation logs.}             
\label{obs_log}      
\centering                          
\begin{threeparttable}
        \begin{tabular}{l c c c c c c c}        
        \hline\hline    \noalign{\smallskip}               
        Target &        Date &  Observing time [UT] & Phase & Exposure time & $N_\mathrm{spectra}$ (A/B) &  Wavelength setting & Seeing \tablefootmark{a} \\     
        \hline  \noalign{\smallskip}                  
  WASP-18b & 2021-10-09 & 04:41-08:32    &  0.28 -- 0.45  & 120 s & 56/56 & K2148 & 0.95$\arcsec$ -- 2.54$\arcsec$ \\ 
  \hline  \noalign{\smallskip}                                
  WASP-76b & 2021-10-31  &  03:57-05:47 & 0.55 -- 0.59  & 120 s & 26/26 & K2148 & 0.55$\arcsec$ -- 0.96$\arcsec$ \\ 
\hline 
        \end{tabular}
\tablefoot{
\tablefoottext{a}{These values are from the header of the FITS file, which was measured at 550\,nm. The seeing in the infrared is expected to be better than these values.}
}
\end{threeparttable}      
\end{table*}

\section{Methods}
We applied the cross-correlation technique \citep{Snellen2010} to search for CO and $\mathrm{H_2O}$ signals in the observed spectra and followed the five procedures listed below, as described in \cite{Yan2022-GIANO}.

\textit{(1) Cleaning up the spectral matrix.} We computed a master spectrum by adding up all the observed spectra and calculated the standard deviation along each pixel column. Then we obtained the empirical S/N of each pixel column by dividing the master spectrum by the standard deviations. Subsequently, we masked the pixel columns that have empirical S/Ns < 15. These masked pixels are mostly bad pixels or pixels located inside strong telluric lines. 
We further performed a three-sigma clip on each spectrum.

\textit{(2) Removing the telluric and stellar lines.} We ran the \texttt{SYSREM} algorithm \citep{Tamuz2005, Birkby2013} in the observer's rest frame to remove the telluric and stellar lines. The \texttt{SYSREM} algorithm was performed on the merged spectrum.
We tested \texttt{SYSREM} iteration numbers of 1$\sim$15 and chose the number that yields the maximum detection significance (see Fig.~\ref{App-sysrem} for CO and Fig.~\ref{App-sysrem-H2O} for $\mathrm{H_2O}$). We find that the best iteration numbers are different for CO and $\mathrm{H_2O}$, which is likely caused by different wavelength ranges as well as different levels of telluric contamination for the CO and $\mathrm{H_2O}$ signals.
The spectra resulting from the removal of telluric and stellar lines, which we refer to as residual spectra, were further shifted to the stellar rest frame. 
Here we adopted stellar systemic velocities of +3.2 km\,s$^{-1}$ for WASP-18 \citep{Hellier2009} and -1.2 km\,s$^{-1}$ for WASP-76 \citep{Ehrenreich2020}. To remove any remaining broadband features, we filtered the residual spectra with a Gaussian high-pass filter that has a standard deviation of 31 points.

\textit{(3) Modeling the thermal emission spectrum.} We modeled the planetary emission spectrum ($F_\text{p}$) using \texttt{petitRADTRANS} \citep{Molliere2019} with the CO line list from \cite{Li2015}. This line list has a temperature range up to 5000\,K. The temperature-pressure ($T$-$P$) profiles are assumed to be two-point parameterized, with the lower pressure point denoted as ($T_1$, $P_1$) and the higher pressure point as ($T_2$, $P_2$) \citep[see][for details]{Yan2020}.
The $T$-$P$ profiles were set to be similar to the ones in \cite{Arcangeli2018} for WASP-18b and \cite{Edwards2020} for WASP-76b. Both of the $T$-$P$ profiles contain temperature inversions. 
We calculated the mixing ratio of CO and $\mathrm{H_2O}$ using the \texttt{easyCHEM} code \citep{Molliere2015} assuming that the atmosphere is under chemical equilibrium conditions with solar metallicity. The obtained model spectra are similar to the model spectra presented in \cite{Yan2022-GIANO}.

\textit{(4) Generating the template grid.} We first added the stellar flux (assumed to be a blackbody spectrum, $F_\text{s}$) to the planetary emission spectrum ($F_\text{p}$). The spectrum of $F_\text{s}$+$F_\text{p}$ mimics the observed spectrum. Since we usually normalize the observed spectrum for high-resolution observations, it is straightforward to express the model as ($F_\text{s}$+$F_\text{p}$)/$F_\text{s}$ = 1 + $F_\text{p}$/$F_\text{s}$. This model spectrum was subsequently convolved with the instrumental profile.
We further created a template grid by shifting the spectrum from --\,500 km\,s$^{-1}$ to +\,500 km\,s$^{-1}$ in 1 km\,s$^{-1}$ steps. The template grid was then sampled into the same wavelength points as the observed spectrum and filtered with the same Gaussian high-pass filter as in step (2). The filtering process acts as a further normalization procedure. 

\textit{(5) Cross-correlating.} We performed the cross-correlation by calculating the weighted cross-correlation function (CCF). The CCF for one spectral frame is
\begin{equation}
\mathrm{CCF_j} = \sum_{i} \frac{r_i \, m_{i,j} }{\sigma_i^2},
\end{equation}
where $r_i$ is the residual spectrum at pixel $i$; $m_{i,j}$ is the template spectrum at pixel $i$ for grid $j$; and $\sigma_i$ is the noise of the observed spectrum that is provided by the pipeline and has been propagated.

\section{Results and discussions}
\subsection{Detection of the CO lines}
We obtained the CCFs of the spectra from nodding position A and position B separately and calculated the corresponding $K_\mathrm{p}$ maps by averaging all the out-of-eclipse CCFs in the planetary rest frame for different $K_\mathrm{p}$ values (0 to 400 km\,s$^{-1}$). Here $K_\mathrm{p}$ is the semi-amplitude of the planetary orbital radial velocity (RV).
We define the x-axis of the $K_\mathrm{p}$ map as $\mathrm{\Delta} \varv$, which means the RV shift relative to the planetary rest frame at a given $K_\mathrm{p}$.
We then added up the $K_\mathrm{p}$ maps from the A and B positions for each planet.
To estimate the detection significance, we measured the noise of the $K_\mathrm{p}$ map by calculating the standard deviation of the regions with $\left| \mathrm{\Delta} \varv \right|$ from 50 to 200 km\,s$^{-1}$ and $K_\mathrm{p}$ from 100 to 300 km\,s$^{-1}$. The $K_\mathrm{p}$ map was then divided by the obtained noise value.

\begin{figure}
   \centering
   \includegraphics[width=0.45\textwidth]{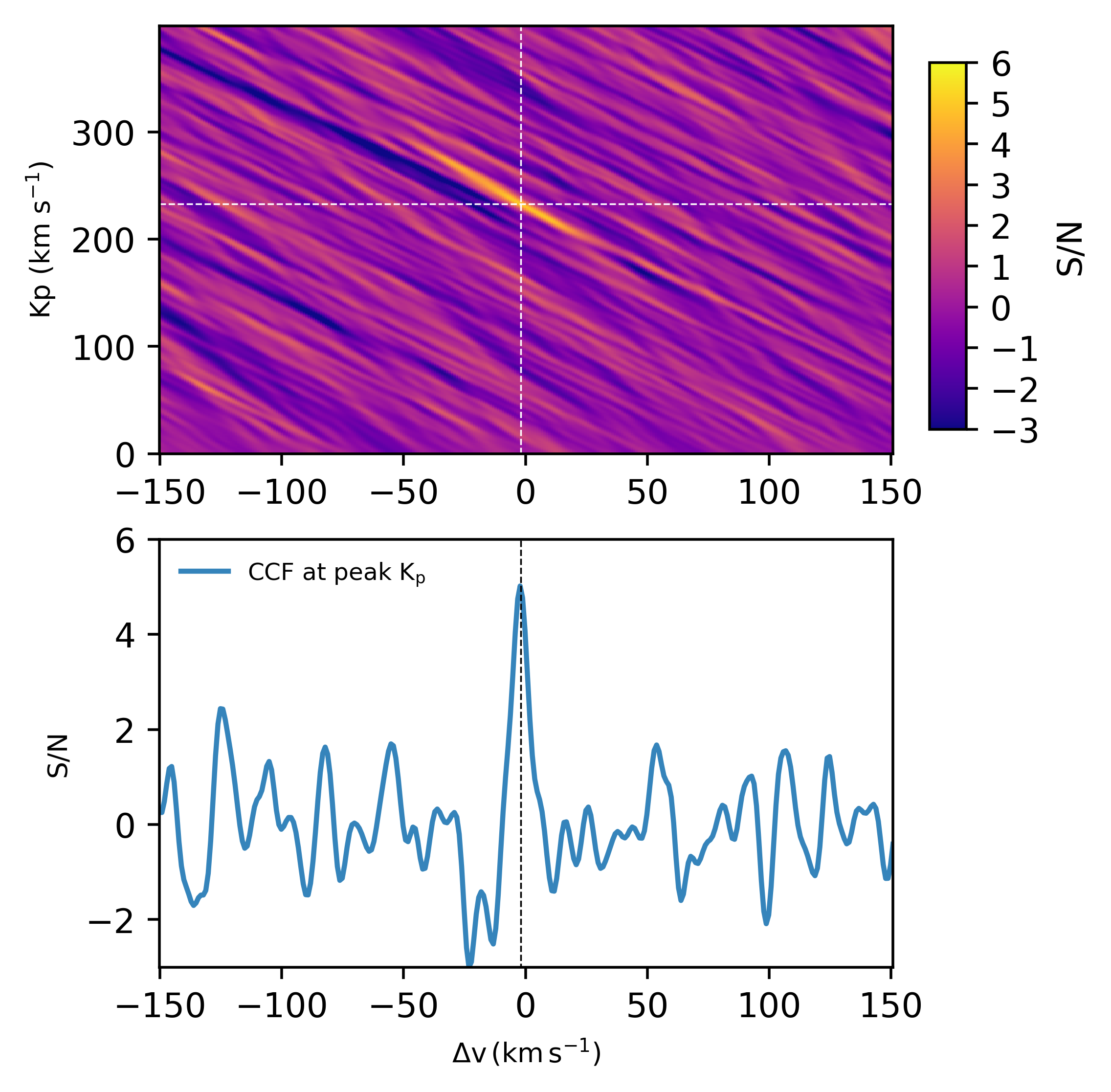}
      \caption{$K_\mathrm{p}$ map (upper panel) and the CCF at maximum S/N (lower panel) for the CO signal of WASP-18b. The crossing of the dashed white lines is the location of the maximum S/N.}
         \label{Kp-W18}
   \end{figure}

\begin{figure}
   \centering
   \includegraphics[width=0.45\textwidth]{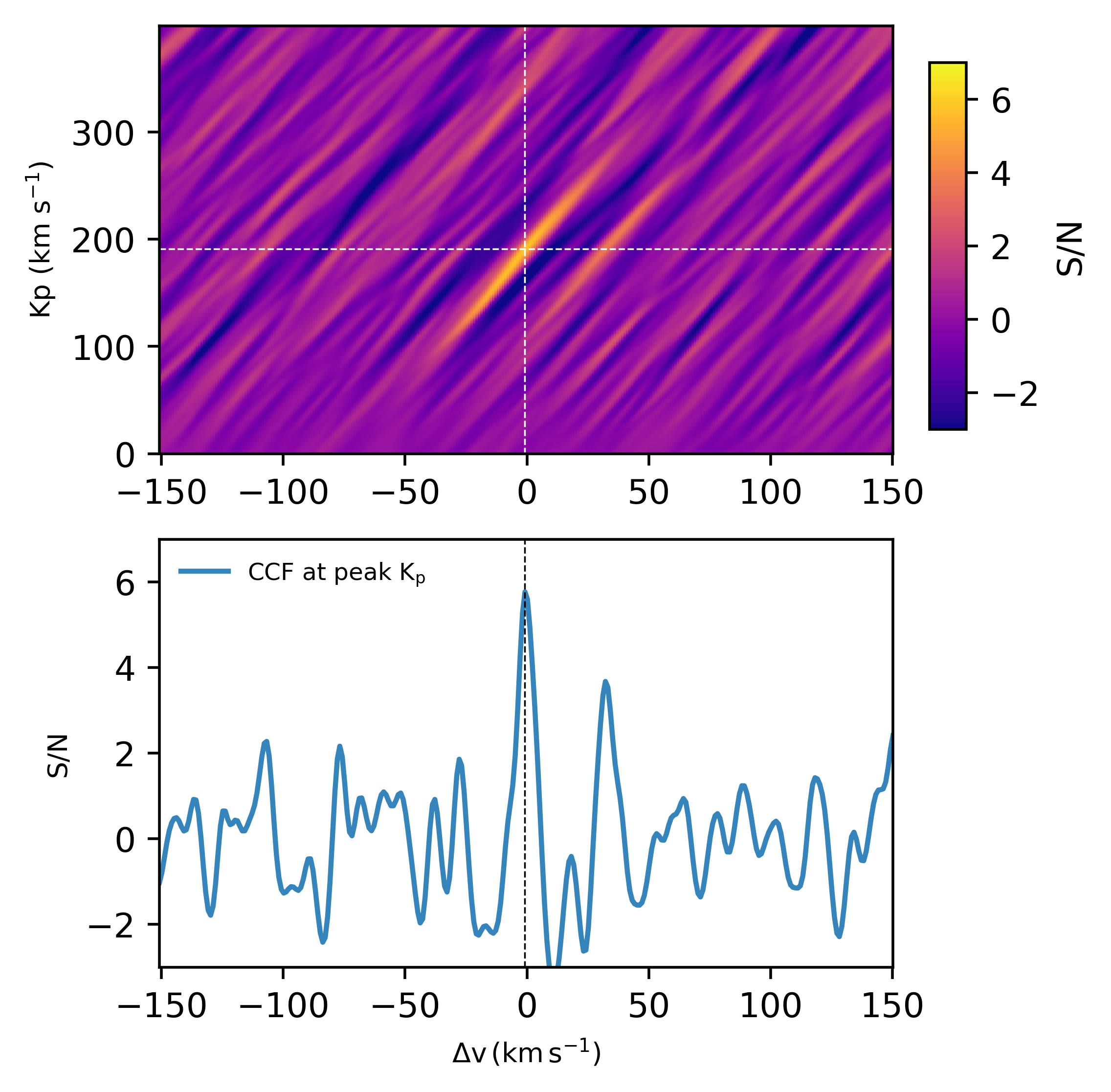}
      \caption{Same as Fig.~\ref{Kp-W18}, but for the CO signal of WASP-76b.}
         \label{Kp-W76}
   \end{figure}

The final results are presented in Figs.~\ref{Kp-W18} and \ref{Kp-W76}. For WASP-18b, we detected the CO signal with a S/N of 5.0 at $K_\mathrm{p} = 233_{-22}^{+42}$ km\,s$^{-1}$ and $\mathrm{\Delta} \varv = -2_{-31}^{+19}$ km\,s$^{-1}$. For WASP-76b, the detected CO signal has a maximum S/N of 5.8 at $K_\mathrm{p} = 191_{-54}^{+31}$ km\,s$^{-1}$ and $\mathrm{\Delta} \varv = -1_{-19}^{+12}$ km\,s$^{-1}$. The location of the maximum S/N is consistent with the expected $K_\mathrm{p}$ value that is inferred from the planetary orbital parameters. For WASP-18b the expected $K_\mathrm{p}$ is $236\pm3$ km\,s$^{-1}$ \citep{Maxted2013}, and for WASP-76b the value is $198\pm1$ km\,s$^{-1}$ \citep{Ehrenreich2020}.
The detection unambiguously confirms that the daysides of the two UHJs have temperature inversion layers, which had previously been inferred from low-resolution emission spectra of \textit{Hubble} and \textit{Spitzer} observations \citep{Arcangeli2018,Edwards2020}.

Temperature inversion layers have also been discovered in several UHJs with high-resolution emission spectroscopy. However, these detections of temperature inversion with CO or \ion{Fe}{i} emission lines are mostly for UHJs orbiting hot A-type stars (Table~\ref{tab-UHJs}). WASP-18b and WASP-76b are the first two such planets orbiting stars with $T_\mathrm{eff}$ < 7000\,K. \cite{Lothringer2019} have predicted that temperature inversion is stronger for UHJs around hotter stars. Therefore, the emission lines from UHJs around hotter stars are stronger and more easily detected.
This may explain why CO and \ion{Fe}{i} emission lines have mostly been discovered in UHJs orbiting A-type stars.

\subsection{Tentative detection of $\mathrm{H_2O}$ emission lines}
\begin{figure}
   \centering
   \includegraphics[width=0.45\textwidth]{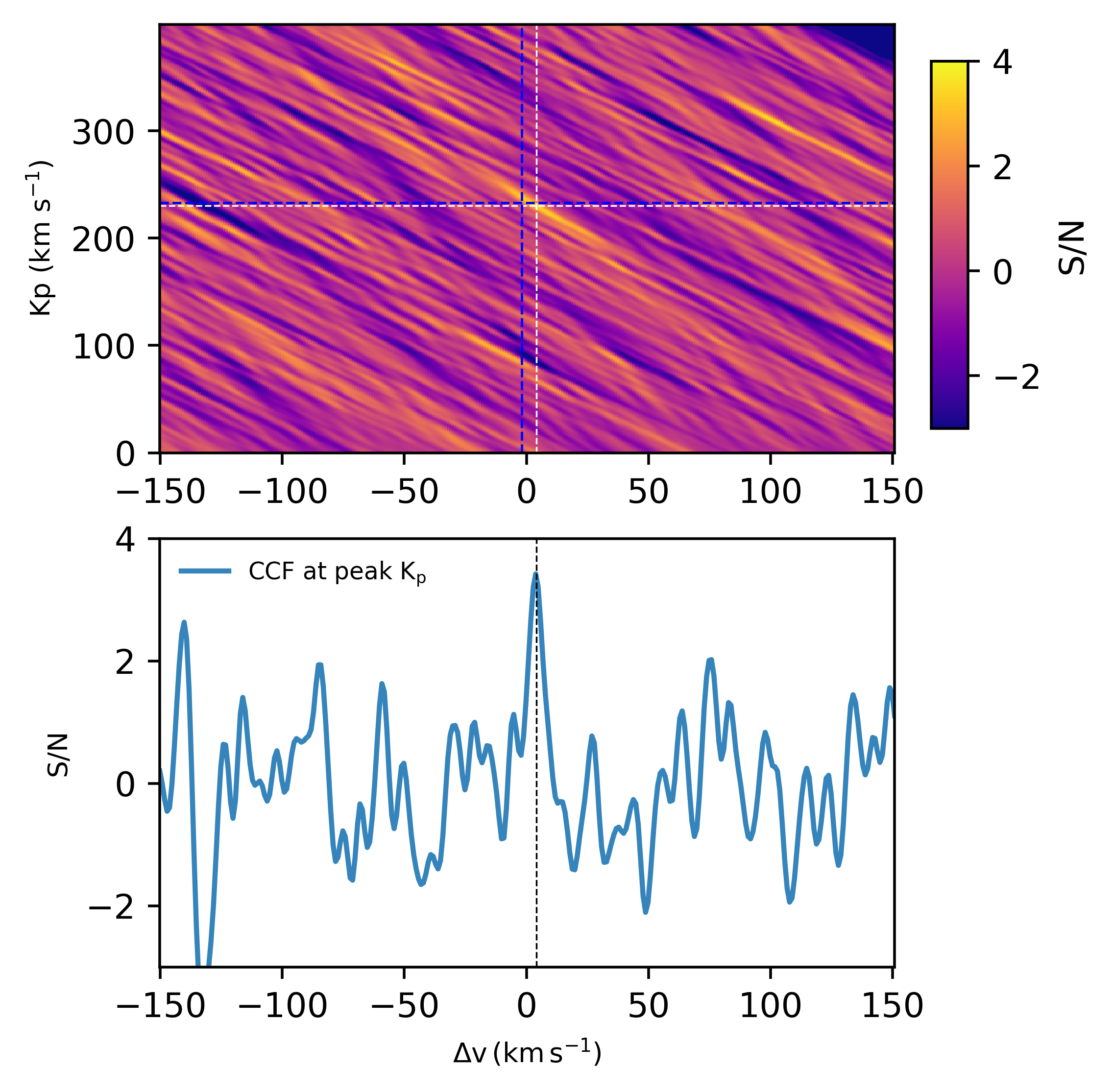}
   \includegraphics[width=0.45\textwidth]{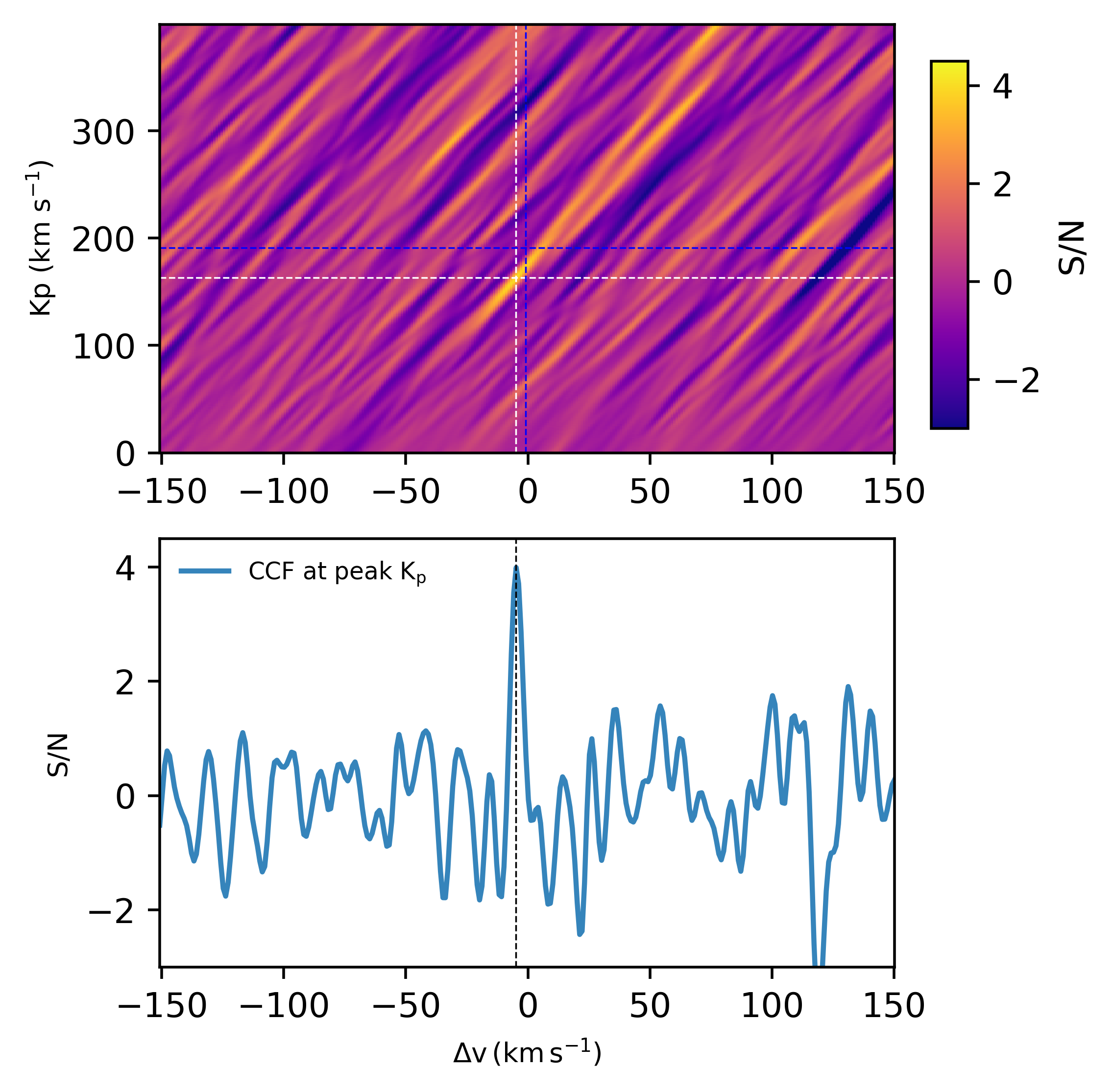}
      \caption{Cross-correlation results of $\mathrm{H_2O}$ emission lines for WASP-18b (upper panel) and WASP-76b (lower panel). The dashed white lines indicate the location of the maximum $\mathrm{H_2O}$ signal.
      The dashed blue lines indicate the location where the CO detection significance is at its maximum.}
         \label{App-H2O}
   \end{figure}

In addition to CO, we also searched for $\mathrm{H_2O}$ lines in the thermal emission spectra. We applied the same method as for the CO detection. The final cross-correlation results are presented in Fig.~\ref{App-H2O}.
We detect tentative $\mathrm{H_2O}$ signals for both planets, with a detection significance of 3$\sim$4 $\sigma$. 
For WASP-18b, the maximum $\mathrm{H_2O}$ signal has a S/N of 3.4 at $K_\mathrm{p} = 230_{-28}^{+19}$ km\,s$^{-1}$ and $\mathrm{\Delta} \varv = 4_{-13}^{+23}$ km\,s$^{-1}$. For WASP-76b, the $\mathrm{H_2O}$ signal has a maximum S/N of 4.0 at $K_\mathrm{p} = 163_{-26}^{+30}$ km\,s$^{-1}$ and $\mathrm{\Delta} \varv = -5_{-10}^{+12}$ km\,s$^{-1}$.
More CRIRES$^+$ observations are needed to confirm the $\mathrm{H_2O}$ detection. 
\cite{Brogi2022} also found evidence of an $\mathrm{H_2O}$ signal in WASP-18b with a significance of 3.3 $\sigma$ using high-resolution spectroscopy. The thermal emission spectra of the two planets have also been observed with the \textit{Hubble} Space Telescope at low resolution, showing non-detections of $\mathrm{H_2O}$ in WASP-18b \citep{Arcangeli2018} and evidence of $\mathrm{H_2O}$ emission in WASP-76b \citep{Edwards2020}.

Unlike CO, $\mathrm{H_2O}$ is expected to exist only at very low altitudes because the species is easily dissociated in the upper layers of UHJ atmospheres (cf. the $\mathrm{H_2O}$ mixing ratios from self-consistent models in Figs.~\ref{Retreival-TP-W18} and \ref{Retreival-TP-W76}). Therefore, CO emission features are expected to be stronger than $\mathrm{H_2O}$ features for UHJs. 

\subsection{Atmospheric retrieval}
\subsubsection{Retrieval framework}
We performed an atmospheric retrieval on the observed CO emission spectra, following the retrieval method described in \cite{Yan2020} and \cite{Yan2022-KELT20b}, which is inspired by \cite{Brogi2019}, \cite{Shulyak2019}, and \cite{Gibson2020}. The retrieval framework used in this work has several changes compared to the previous method in \cite{Yan2020}, including  the application of \texttt{SYSREM} filtering and the introduction of rotational broadening. Furthermore, the retrieval was performed on the matrix of the residual spectra (i.e., the spectral matrix after the \texttt{SYSREM} processing) instead of on the combined master residual spectrum as in \cite{Yan2020}. Below we describe the details of the retrieval framework.

We first calculated forward models with \texttt{petitRADTRANS} assuming a two-point $T$-$P$ profile. We assumed that the volume mixing ratio of CO (log\,(CO)) is constant along the pressure axis and set it as a free parameter. The mean molecular weight was set to vary with the $T$-$P$ profile, and its value was calculated using the \texttt{easyCHEM} code assuming solar abundance. 
We then generated a model spectral matrix that has the same dimension as the residual spectra. Since the residual spectra are in the observer's rest frame, we shifted each model spectrum with the RV of the corresponding planetary orbital velocity and the systemic velocity. In addition, we added an RV shift of $\mathrm{\Delta} \varv$ to the model spectrum to account for any additional RV deviation from the planetary rest frame. 
We subsequently broadened the model spectrum and sampled the spectrum into the wavelength grid of the observed spectra.

\begin{figure}
   \centering
   \includegraphics[width=0.45\textwidth]{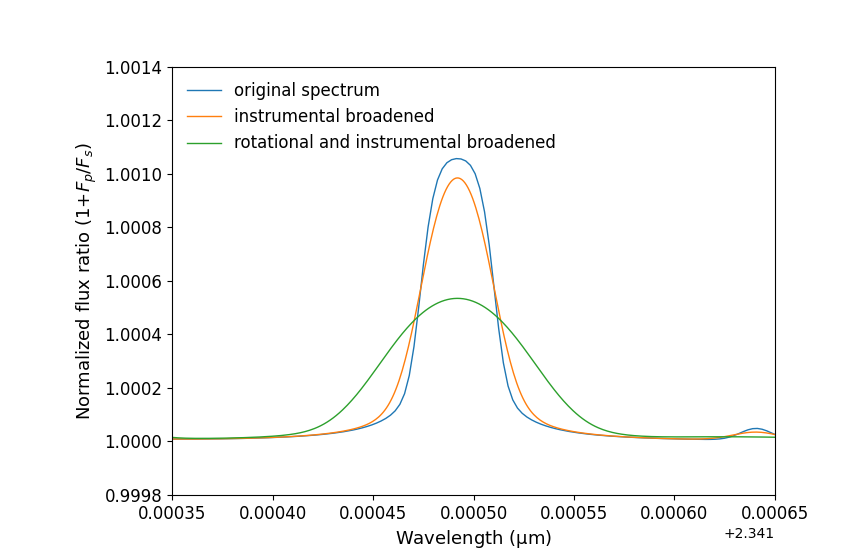}
      \caption{Illustration of the line profile broadening. The spectrum is the modeled CO emission line for WASP-18b. }
         \label{line-profile}
   \end{figure}

For the retrieval of high-resolution observations, the line profile broadening is important (Fig.~\ref{line-profile}). The pressure and thermal broadening was already taken into account during the computation of the opacity grid. We also convolved the model spectra with the instrumental profile (assumed to be a Gaussian profile corresponding to \textit{R} = 120\,000 for WASP-18b and 130\,000 for WASP-76b). Here, we introduced an additional parameter, $\varv_\mathrm{eq}$ (the equatorial rotation velocity), to account for the rotational broadening. We followed Eq.~(3) in \cite{Diaz2011} to calculate the rotational profile. The formula is originally from \cite{Gray1992} and has been widely used for studying stellar rotation.
We assumed a linear limb-darkening law and fixed the coefficient $\epsilon$ to 1, which means the limb part of the hemisphere has no contribution to the total flux. Unlike the stellar atmosphere, the temperature distribution of the planetary atmosphere is extremely inhomogeneous. Therefore, the coefficient $\epsilon$  here represents both the limb darkening and the inhomogeneous temperature distribution.
Since the planet is likely tidally locked, we assumed the inclination angle of the planetary equator to be the same as the planetary orbital inclination angle, leading to $\mathrm{sin}\,i \approx 1$. We then convolved the model spectrum with this rotational profile in velocity space (i.e., ln~$\lambda$).

\begin{figure}
   \centering
   \includegraphics[width=0.48\textwidth]{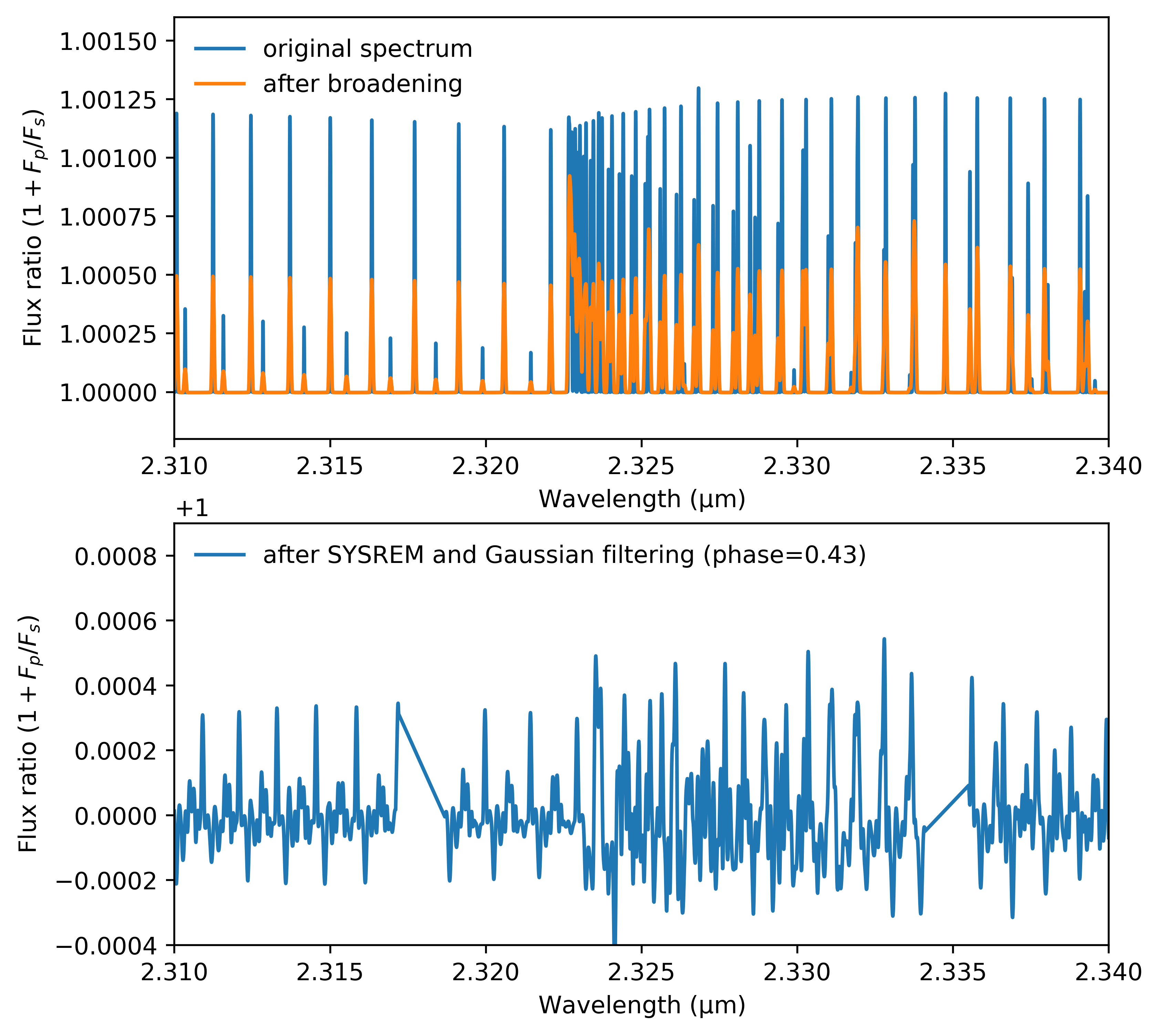}
      \caption{Example of model processing for the atmospheric retrieval. Upper panel: Original spectrum and the spectrum after broadening for WASP-18b. Lower panel: Spectrum after applying the \texttt{SYSREM} filtering and the Gaussian high-pass filtering. Here the spectrum is the model spectrum at phase 0.43 and is sampled into the CRIRES$^+$ wavelength points.}
         \label{model-process}
   \end{figure}

Before fitting the model matrix with the residual spectra, proper filtering of the model is required. During the data reduction, we performed the \texttt{SYSREM} algorithm on the data, which altered the strength and profile of the planetary lines. The actual distortion of the line profile is also phase-dependent. Therefore, the model spectral matrix also had to be processed with the \texttt{SYSREM} algorithm. However, running the \texttt{SYSREM} algorithm on each model spectrum during the retrieval is time-consuming. Here we applied the fast \texttt{SYSREM} filtering technique as described by \cite{Gibson2022}. This technique boosts the \texttt{SYSREM} processing on the model matrix by using information from the \texttt{SYSREM} calculation of the data matrix.
After the \texttt{SYSREM} filtering of the model matrix, we also applied the Gaussian high-pass filtering to the model spectra in a similar way as performed on the observed spectra. An example of the model processing is presented in Fig.~\ref{model-process}.

The retrieval was achieved by evaluating the likelihood function with \texttt{emcee} \citep{Mackey2013}, which conducts Markov chain Monte Carlo (MCMC) simulations. The logarithm of the likelihood function ($L$) is expressed as  \citep{Hogg2010}\footnote{https://emcee.readthedocs.io/en/stable/tutorials/line/}
\begin{equation}
      \mathrm{ln}(L) = -\frac{1}{2}\sum_{i,j} \left[ \frac{(R_{i,j} - M_{i,j})^2}{(\beta \sigma_{i,j})^2} + \mathrm{ln}(2 \pi (\beta \sigma_{i,j})^2) \right] ,
\label{eq-mcmc}
\end{equation}
where $R_{i,j}$ is the matrix of the residual spectra at wavelength point $i$ and time $j$; $M_{i,j}$ is the matrix of the model spectra; and $\beta$ is a scaling factor to the noise, $\sigma_{i,j}$.
The free parameters and their boundaries are listed in Table \ref{tab-mcmc}. We assumed uniform priors for all the parameters.
We ran the MCMC simulation with 12000 steps and 24 walkers for each free parameter.

   \begin{figure*}
   \centering
   \includegraphics[width=0.42\textwidth]{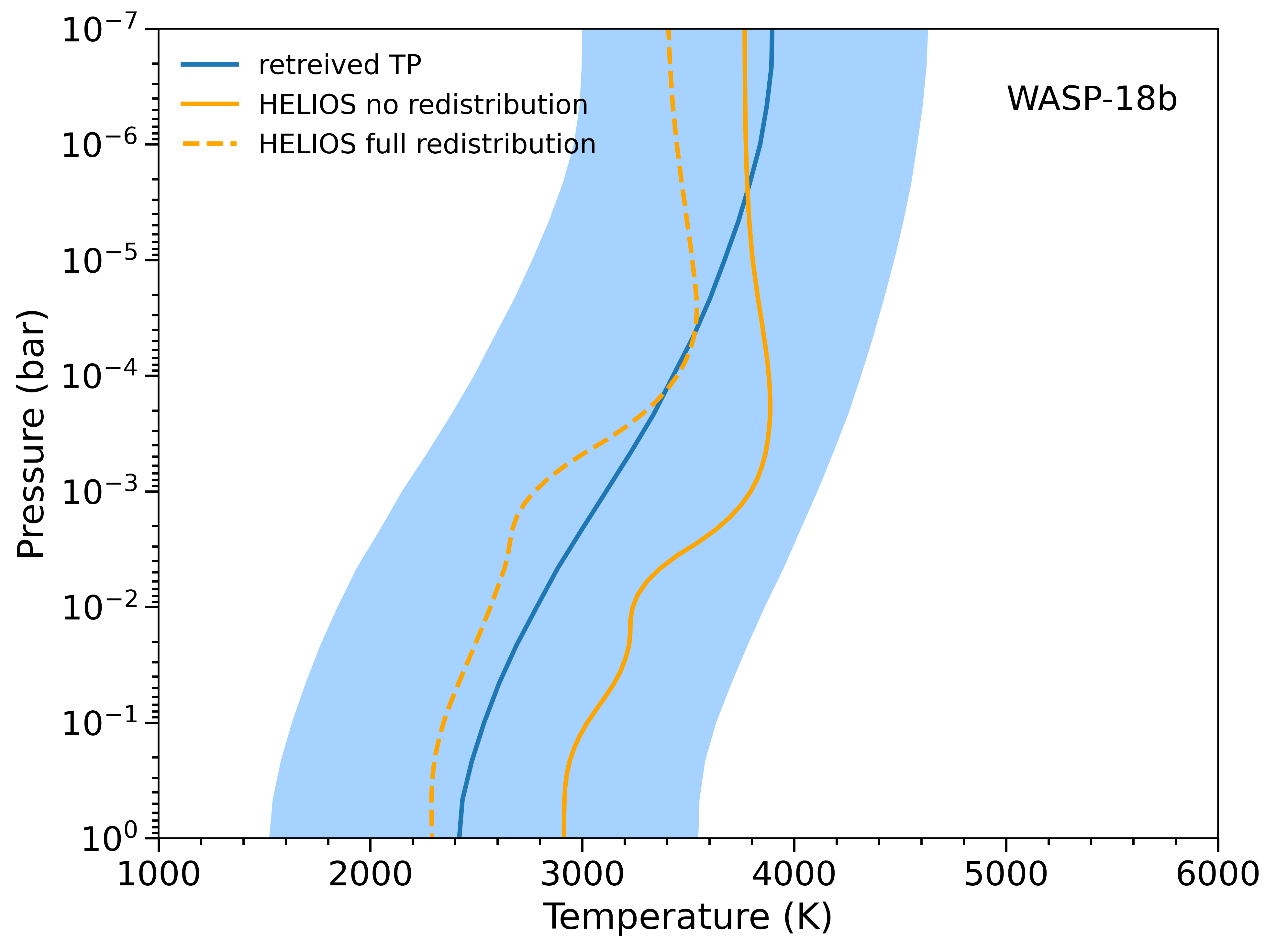}
   \includegraphics[width=0.42\textwidth]{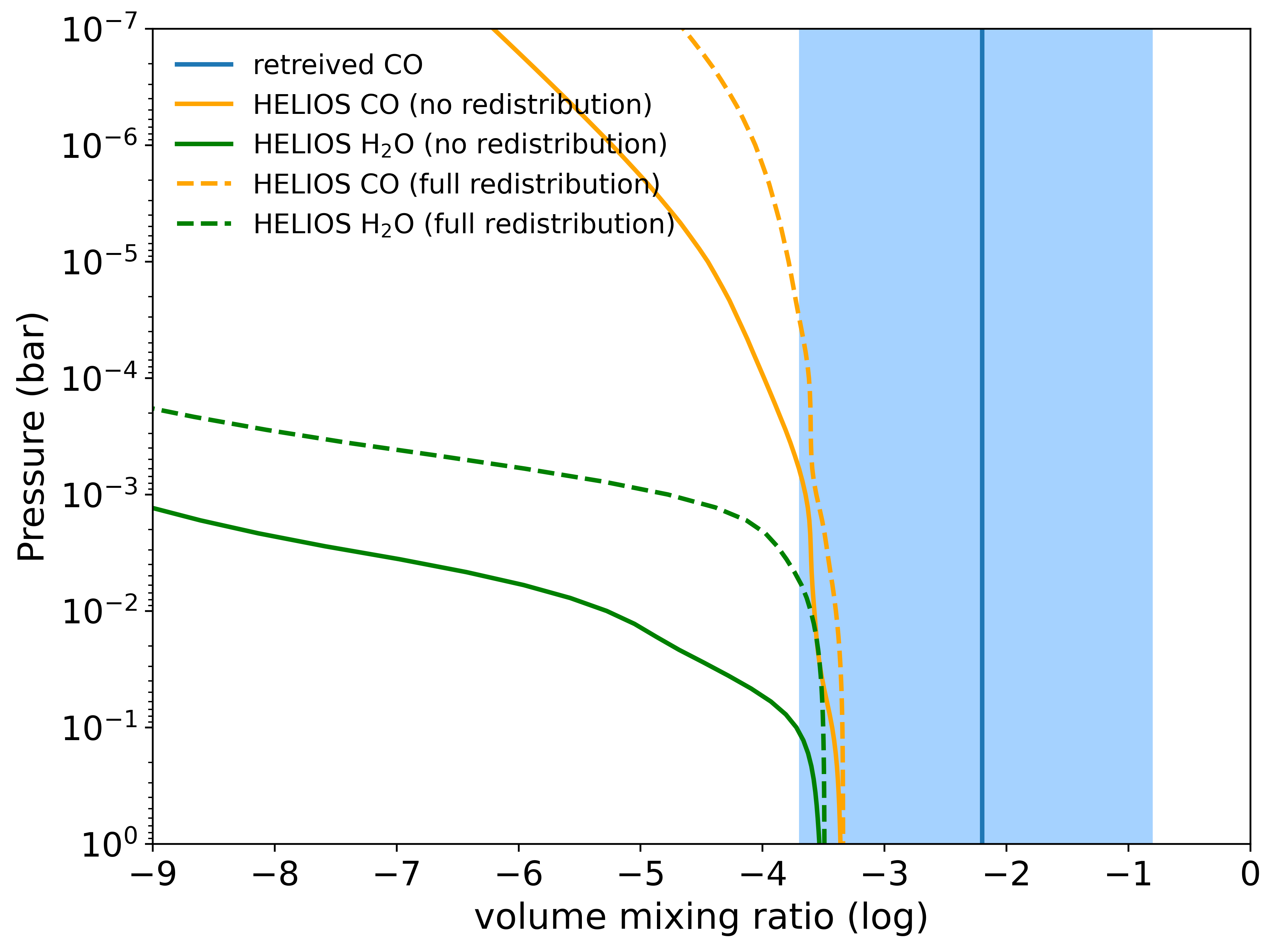}
      \caption{Retrieved $T$-$P$ profile (left panel) and CO mixing ratio (right panel) for WASP-18b and comparison with the results from the self-consistent \texttt{HELIOS} model. The \texttt{HELIOS} model is calculated assuming solar abundances for both no and full heat redistribution from dayside to nightside. The blue shadows are the 1$\mathrm{\sigma}$ range of the retrieved results. }
         \label{Retreival-TP-W18}
   \end{figure*}

   \begin{figure*}
   \centering
   \includegraphics[width=0.42\textwidth]{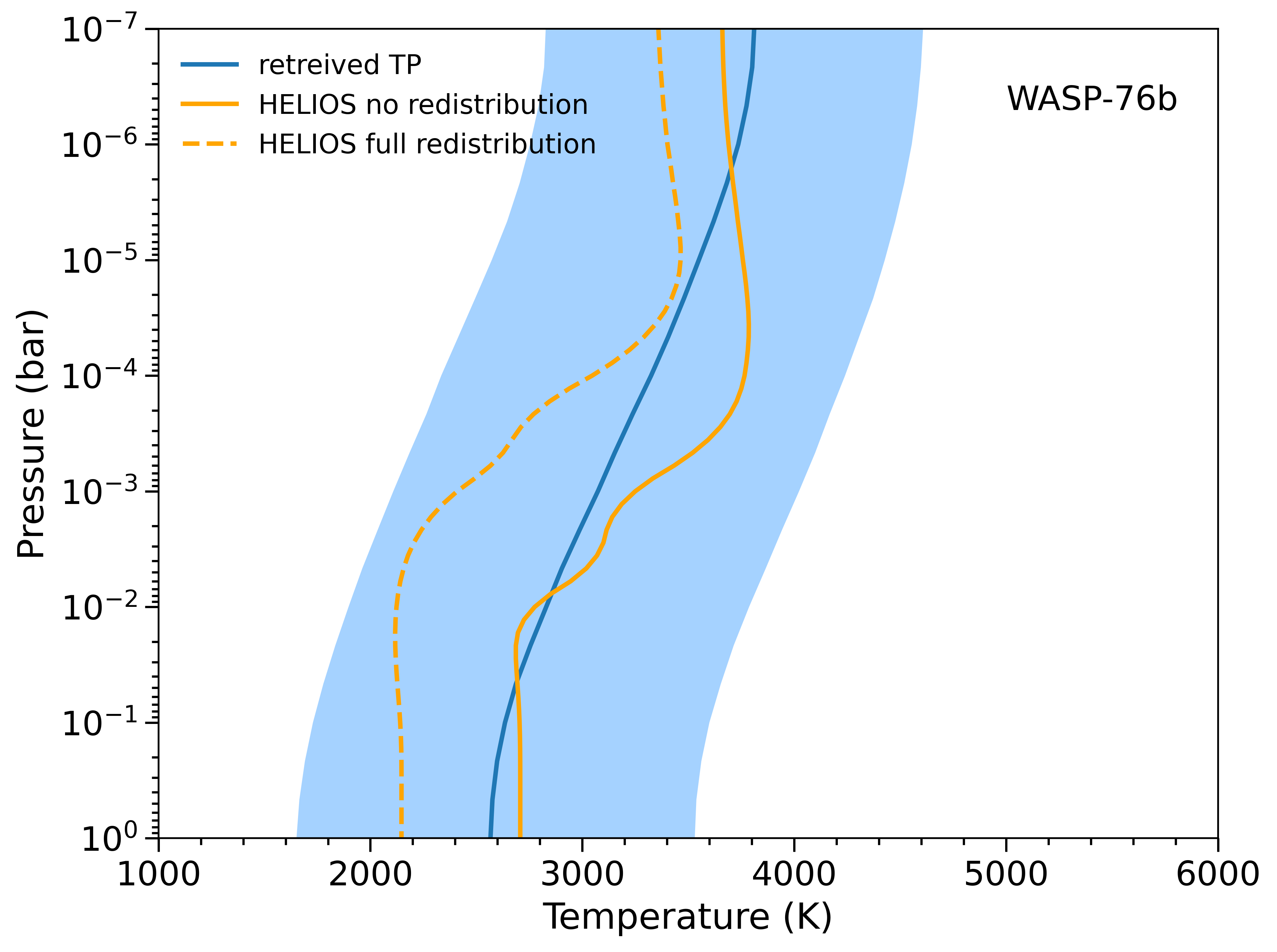}
   \includegraphics[width=0.42\textwidth]{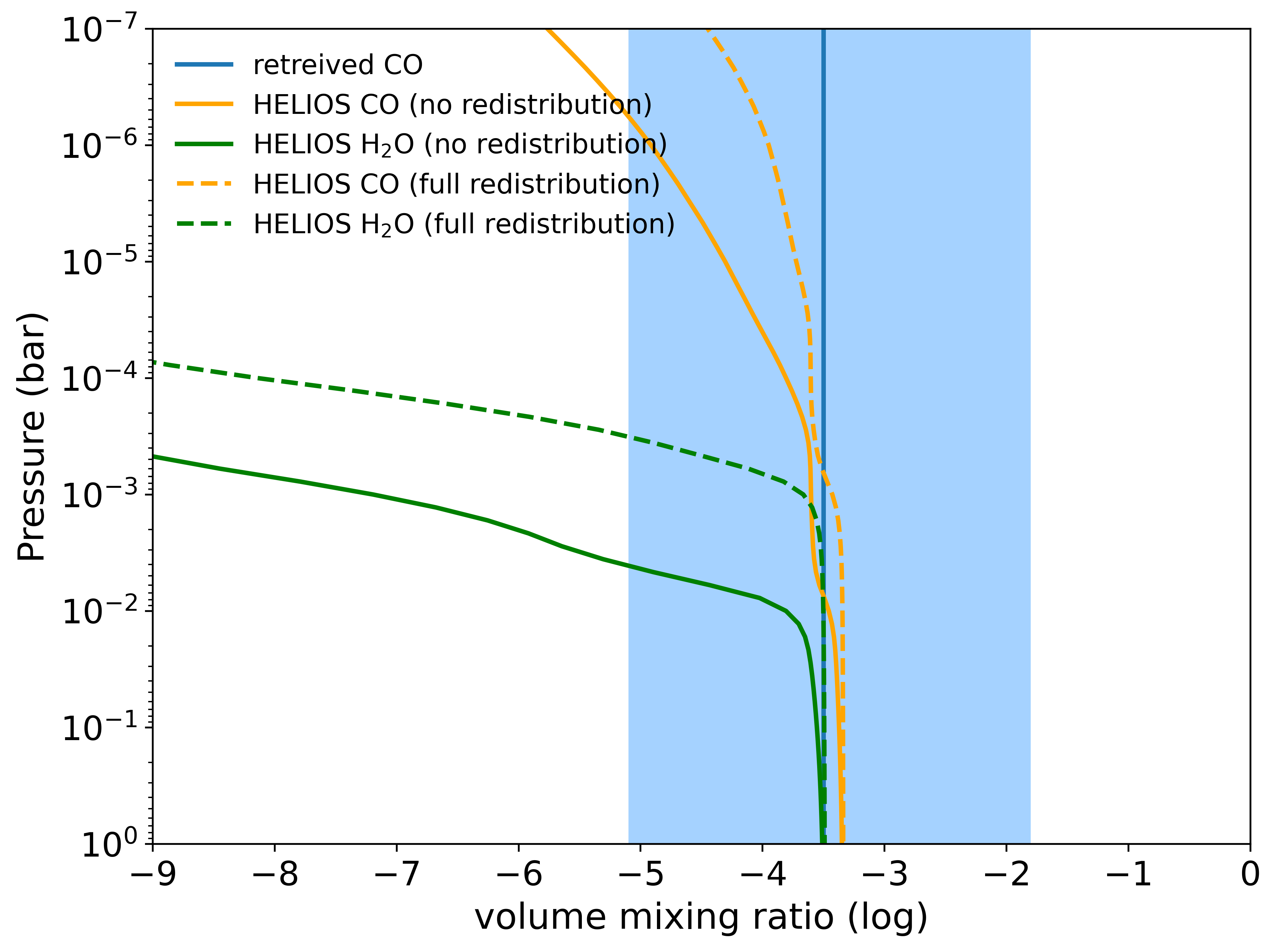}
      \caption{Same as Fig.~\ref{Retreival-TP-W18}, but for WASP-76b. }
         \label{Retreival-TP-W76}
   \end{figure*}

\subsubsection{Retrieval results}
The retrieved $T$-$P$ profile is presented in Figs.~\ref{Retreival-TP-W18} and \ref{Retreival-TP-W76}, and the values of the parameters are presented in Table \ref{tab-mcmc}, with their posterior distribution shown in Figs.~\ref{App-corner-W18} and \ref{App-corner-W76}.
For comparison, we calculated self-consistent models using the modified \texttt{HELIOS} code \citep{Malik2017}, which includes opacities due to neutral and singly ionized species \citep{Fossati2021-NLTE}. Here, we assumed an atmosphere with solar abundance. For the heat redistribution from dayside to nightside, we assumed two extreme cases (i.e., no and full heat redistribution). The mixing ratios of the chemical species were calculated with \texttt{FastChem} \citep{Stock2018}.
A detailed description of the \texttt{HELIOS} model calculation can be found in \cite{Yan2022-KELT20b}.

For both planets, the retrieved $T$-$P$ profile clearly shows the existence of temperature inversion, which is in alignment with the predictions from the \texttt{HELIOS} model. In general, the inversions of both planets are significant but weaker than the inversions of UHJs orbiting A-type stars (e.g., WASP-189b and KELT-20b). In addition, the inversion of WASP-18b is located at lower altitudes than that of WASP-76b from the \texttt{HELIOS} models, which is due to the large surface gravity of WASP-18b.

The retrieved CO volume mixing ratio is $-2.2_{-1.5}^{+1.4}$ for WASP-18b and $-3.6_{-1.6}^{+1.8}$ for WASP-76b. 
We set log~(CO) as a constant during the retrieval, though the actual mixing ratio becomes smaller at higher altitudes due to thermal dissociation. Therefore, the retrieved log~(CO) should be regarded as the average value in the inversion layer.
The theoretical CO mixing ratio under the solar metallicity assumption is around --3.5 (cf. the \texttt{HELIOS} result in Figs.~\ref{Retreival-TP-W18} and \ref{Retreival-TP-W76}). Therefore, the retrieved log~(CO) of WASP-18b is noticeably higher but within error bars when compared to the solar abundance value, which could be due to a higher metallicity or a higher C/O ratio. This agrees with the results from \cite{Arcangeli2018} and \cite{ Brogi2022}. However, we emphasize that the deviation from the solar abundance is only one $\sigma$.
The retrieved log~(CO) of WASP-76b broadly agrees with the solar abundance value.

Similar to previous studies \citep[e.g.,][]{Yan2020}, we find there is some degeneracy between the log~(CO) and the altitude of the inversion (i.e., $P_1$ and $P_2$), which can be identified in the corner plot in Figs.~\ref{App-corner-W18} and \ref{App-corner-W76}. More observations with higher data quality in combination with low-resolution spectra should be able to better constrain the CO mixing ratio and the location of the inversion.

The observed CO emission lines have a relatively broad profile. The retrieved equatorial velocity, $\varv_\mathrm{eq}$, is $7.0\pm{2.9}$ km\,s$^{-1}$ for WASP-18b and $5.2_{-3.0}^{+2.5}$ km\,s$^{-1}$ for WASP-76b. An example of the best-fit line profile for WASP-18b is presented as the green line in Fig.~\ref{line-profile}. 
When assuming a tidally locked rotation, we obtain an equatorial velocity of 6.65 km\,s$^{-1}$ for WASP-18b and 5.26 km\,s$^{-1}$ for WASP-76b. Therefore, the retrieved rotational velocities agree well with the tidally locked values. 

The retrieved $\mathrm{\Delta} \varv$ is $-4.3_{-0.9}^{+1.0}$ km\,s$^{-1}$ for WASP-18b and $2.1_{-0.7}^{+0.8}$ km\,s$^{-1}$ for WASP-76b. The values imply that the CO signals deviate from the planetary rest frame, which is likely due to the combination of planetary rotation and atmospheric circulation. For example, the WASP-76b observation was performed at orbital phases right after the eclipse (0.55 -- 0.59); therefore, a dayside to nightside wind could result in a redshifted CO signal. For WASP-18b, the observation covers a large range of orbital phases before the eclipse (0.28 -- 0.45). The blueshifted $\mathrm{\Delta} \varv$ of the WASP-18b signal is probably driven by the spectra taken at the beginning of the observation (i.e., close to the quadrature) since the flux drops dramatically during the observation. Therefore, either planetary rotation or day-to-night wind could yield a blueshifted signal around phase $\sim$ 0.3 for WASP-18b.
The actual atmospheric circulation patterns from the general circulation models are rather complicated \citep[e.g.,][]{Tan2019}, and further phase-resolved observations with high S/Ns are required to study the detailed circulation pattern. Furthermore, the retrieved $\mathrm{\Delta} \varv$ also depends on the $K_\mathrm{p}$ value, which was fixed in the above retrieval. When considering the uncertainties of $K_\mathrm{p}$ inferred from planetary orbital parameters (i.e., $236\pm3$ km\,s$^{-1}$ for WASP-18b and $198\pm1$ km\,s$^{-1}$ for WASP-76b), the uncertainty of the retrieved $\mathrm{\Delta} \varv$ increases by $\sim$ 2 km\,s$^{-1}$ for WASP-18b and $\sim$ 0.5 km\,s$^{-1}$ for WASP-76b. More observations that cover orbital phases both before and after eclipse will enable us to better constrain  $K_\mathrm{p}$ and reduce the uncertainty of $\mathrm{\Delta} \varv$.

%
\begin{table}
\small 
\caption{Retrieved values from the CO emission spectra.}             
\label{tab-mcmc}      
\centering                          
\begin{threeparttable}
        \begin{tabular}{l c c c}        
        \hline\hline \noalign{\smallskip}                 
                Parameter & WASP-18b & WASP-76b  &  Boundaries [Unit] \\     
        \hline     \noalign{\smallskip}                   
  \rule{0pt}{2.5ex} log (CO) & $-2.2_{-1.5}^{+1.4}$ & $-3.6_{-1.6}^{+1.8}$ &  -10 to 0  \\ 
    \rule{0pt}{2.5ex} $T_\mathrm{1}$ & $3900_{-900}^{+700}$ & $3800_{-1000}^{+800}$ &  1000 to 5000 [K] \\  
  \rule{0pt}{2.5ex} log $P_\mathrm{1}$ & $-4.9_{-1.5}^{+2.0}$ & $-5.2_{-1.3}^{+2.0}$ & $-7$ to 0 [log bar] \\    
  \rule{0pt}{2.5ex} $T_\mathrm{2}$ & $2400_{-900}^{+1100}$ & $2600_{-900}^{+1000}$ & 1000 to 5000 [K] \\ 
  \rule{0pt}{2.5ex} log $P_\mathrm{2}$ & $-1.4_{-1.6}^{+1.0}$ & $-1.6_{-2.0}^{+1.2}$ & $-7$ to 0 [log bar] \\  
  \rule{0pt}{2.5ex} $K_\mathrm{p}$ & 236 (fixed) & 198 (fixed) &  [km\,s$^{-1}$] \\  
  \rule{0pt}{2.5ex} $\mathrm{\Delta} \varv$ & $-4.3_{-0.9}^{+1.0}$ & $2.1_{-0.7}^{+0.8}$ & -20 to 20 [km\,s$^{-1}$] \\ 
    \rule{0pt}{2.5ex} $\varv_\mathrm{eq}$ & $7.0\pm{2.9}$ & $5.2_{-3.0}^{+2.5}$ & 0 to 20 [km\,s$^{-1}$] \\               
\noalign{\smallskip}  \hline                                   
        \end{tabular}
\end{threeparttable}      
\end{table}

\section{Conclusions}
We observed the thermal emission spectra of WASP-18b and WASP-76b using the guaranteed time of the CRIRES$^+$ consortium. By applying the cross-correlation method, we detected strong CO emission lines and found evidence of $\mathrm{H_2O}$ signals, which indicates the existence of temperature inversion layers in the two UHJs. 
We further performed forward-model retrievals with the observed CO lines. The retrieval indicates that the CO volume mixing ratio of WASP-18b is slightly higher than the theoretical value computed under the solar metallicity assumption. The observed CO line profiles are relatively broad compared to the instrumental broadening. We further introduced rotational broadening and retrieved the equatorial rotation velocity, $\varv_\mathrm{eq}$ ($7.0\pm{2.9}$ km\,s$^{-1}$ for WASP-18b and $5.2_{-3.0}^{+2.5}$ for WASP-76). These $\varv_\mathrm{eq}$ values agree with the tidally locked rotation velocities.

CO emission lines have previously been discovered in three UHJs orbiting A-type stars. All the previous discoveries of Fe emission lines have also been in UHJs orbiting A-type stars. The host stars of WASP-18b and WASP-76b are F-type, making them the first UHJs with emission lines detected while orbiting stars with $T_\mathrm{eff}$ < 7000\,K. Future observations targeting UHJs around even cooler stars will provide more observational evidence on how the temperature and chemical structures of UHJs change with stellar type.

CRIRES$^+$ is an ideal instrument for exoplanet atmosphere observations because of its high resolution, high stability, and relatively large wavelength coverage in the near-infrared. There is also a great potential to perform a combined analysis of high-resolution observations from CRIRES$^+$ with low-resolution observations from the \textit{James Webb} Space Telescope.

\begin{acknowledgements}
We thank the anonymous referee for the useful comments.
CRIRES$^+$ is an ESO upgrade project carried out by Th\"uringer Landessternwarte Tautenburg, Georg-August Universit\"at G\"ottingen, and Uppsala University. The project is funded by the Federal Ministry of Education and Research (Germany) through Grants 05A11MG3, 05A14MG4, 05A17MG2 and the Knut and Alice Wallenberg Foundation.
Based on observations collected at the European Organisation for Astronomical Research in the Southern Hemisphere under ESO programmes 108.22PH.001 and 108.22PH.002.
F.Y. acknowledges the support of Frontier Scientific Research Program of Deep Space Exploration Laboratory (2022-QYKYJH-ZYTS-016).
D.S. acknowledges the financial support from the State Agency for Research of the Spanish MCIU through the ``Center of
Excellence Severo Ochoa'' award to the Instituto de Astrof\'isica de Andaluc\'ia (SEV-2017-0709).
M.R. acknowledges the support by the DFG priority program SPP 1992 ``Exploring the Diversity of Extrasolar Planets'' (DFG PR 36 24602/41).
\end{acknowledgements}

\bibliographystyle{aa} 

\bibliography{CRIRES-CO-refer}

\begin{appendix}
\section{Additional tables and figures}


   \begin{figure}
   \centering
   \includegraphics[width=0.45\textwidth]{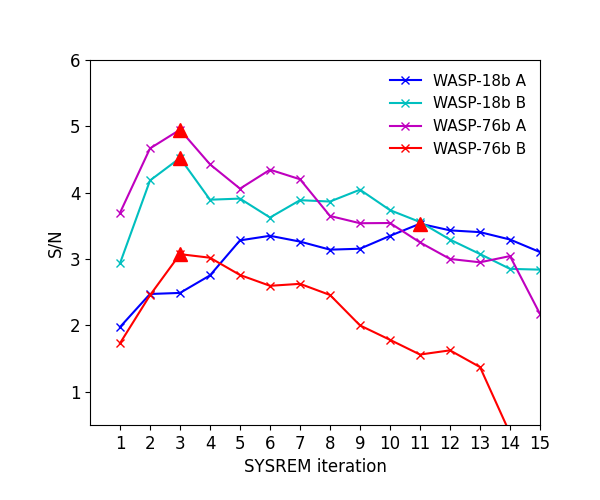}
      \caption{Detection significance of the CO signal with different \texttt{SYSREM} iteration numbers. The maximum S/N values are marked with the red triangles. The value here is measured at fixed $K_\mathrm{p}$ and at the $\mathrm{\Delta} \varv$ location where the detection signal is the strongest.}
         \label{App-sysrem}
   \end{figure}

   \begin{figure}
   \centering
   \includegraphics[width=0.45\textwidth]{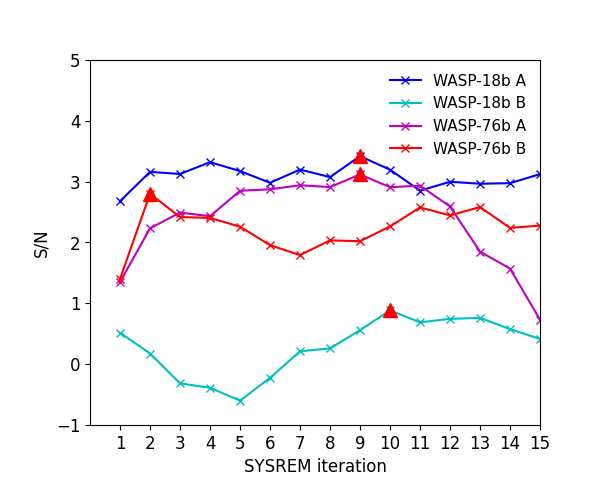}
      \caption{Same as Fig.~\ref{App-sysrem}, but for the $\mathrm{H_2O}$ signal.}
         \label{App-sysrem-H2O}
   \end{figure}

   \begin{figure*}
   \centering
   \includegraphics[width=0.7\textwidth]{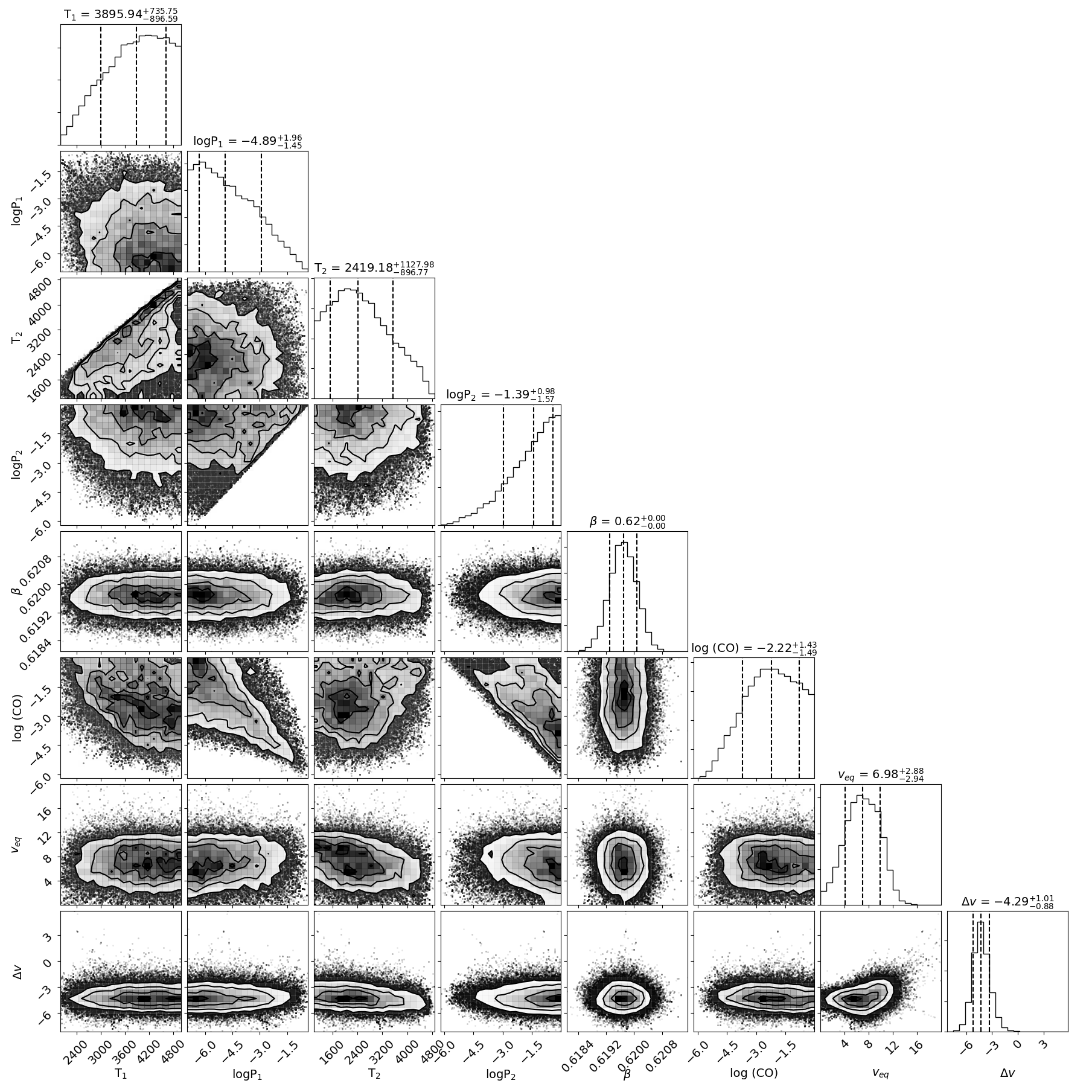}
      \caption{Posterior distribution of the parameters from the retrieval of CO emission lines for WASP-18b.}
         \label{App-corner-W18}
   \end{figure*}

   \begin{figure*}
   \centering
   \includegraphics[width=0.7\textwidth]{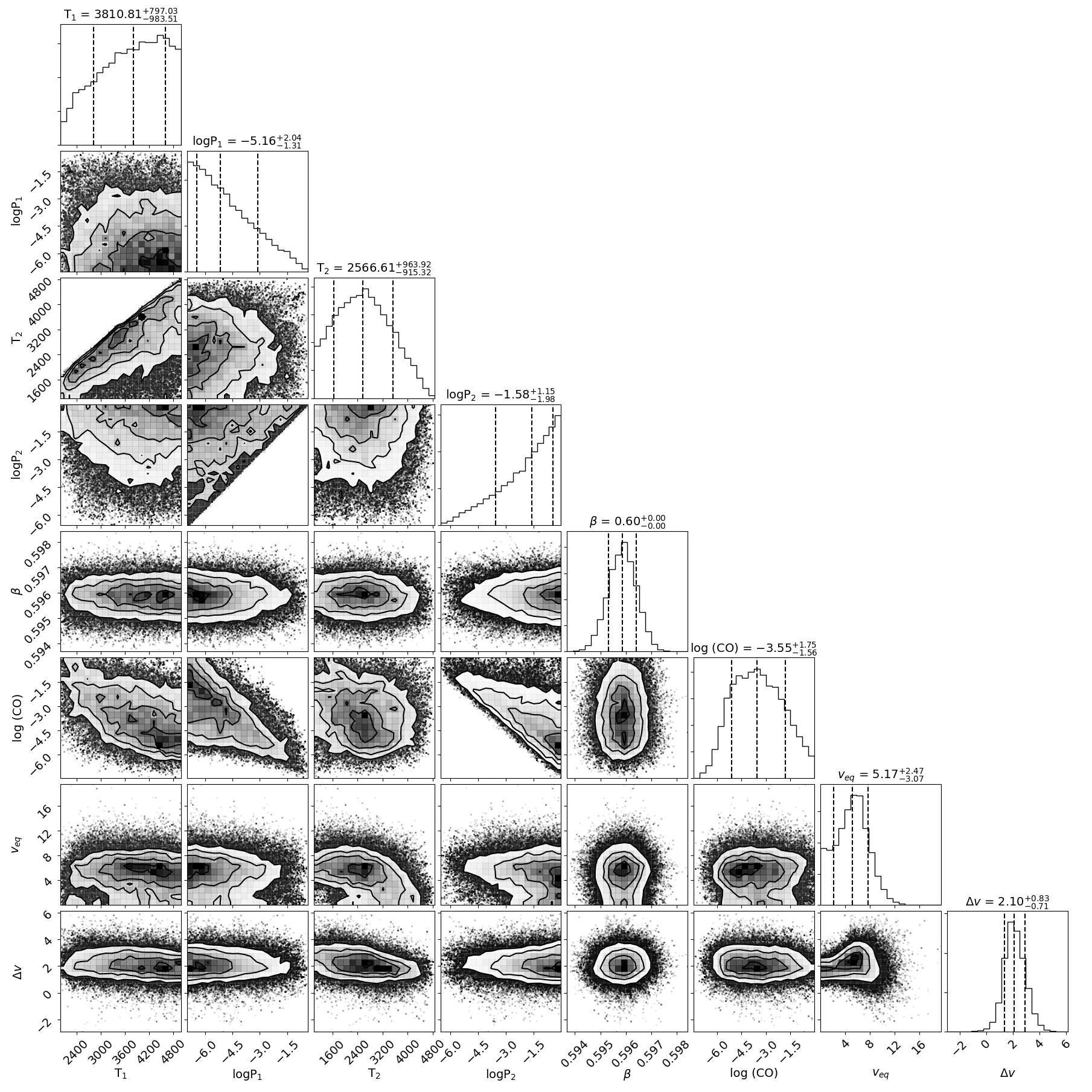}
      \caption{Same as Fig.~\ref{App-corner-W18}, but for WASP-76b.}
         \label{App-corner-W76}
   \end{figure*}

\end{appendix}

\end{document}